\documentclass[useAMS,usenatbib]{mn2e}
\usepackage{graphicx}

\title[Energy Ranking Preservation]{Energy Ranking Preservation in a N-Body Cosmological Simulation}
\author[C. C. Dantas and F. M. Ramos]{Christine C. Dantas$^{1}$\thanks{E-mail:
ccdantas@iae.cta.br} and Fernando M. Ramos$^{2}$\thanks{E-mail: fernando@lac.inpe.br}\\
$^{1}$Instituto de Aeron\'autica e Espa\c co (IAE/CTA), P\c ca. Mal. Eduardo Gomes, 50, Vila das Ac\'acias, 12228-904,\\ S\~ao Jos\'e dos Campos, SP, Brazil\\
$^{2}$Instituto Nacional de Pesquisas Espaciais (INPE), Caixa Postal 515,
12201-970, S\~ao Jos\'e dos Campos, SP, Brazil }

\begin{document}

\date{in original form: \today}

\maketitle

\label{firstpage}

\begin{abstract}
In this paper we present a study of the cosmic flow from
the point of view of how clusterings at different dynamical regimes in
an expanding universe evolve according to a `coarse-grained'
partitioning of their ranked energy distribution.   By analysing
a $\Lambda$-CDM cosmological simulation from the Virgo Project, we
find that cosmic flows evolve in an orderly sense, 
when tracked from their coarse-grained energy cells,  even when 
nonlinearities are already developed.  
We show that it is possible to characterize scaling laws
for the Pairwise Velocity Distribution in terms of the energy cells, 
generally valid at the linear and nonlinear clustering regimes.

\end{abstract}


\section{Introduction}

Cosmological N-body simulations play a crucial role in
the study of the formation of cosmic structure. These structures
are believed to have grown from weak density fluctuations present 
in the otherwise homogeneous and rapidly expanding early universe. 
These fluctuations were amplified by gravity, eventually turning into 
the rich structure that we see around us today \citep{spr05}. Cosmological 
N-body simulations usually start from initial conditions set at some early epoch, 
then follow the later evolution of the dark matter (and, if included,
other matter components) into the nonlinear regime
where it can be compared with the large-scale structure in
galaxy surveys. In these simulations, the dominant mass component, 
the cold dark matter, is assumed to be made of elementary particles that 
interact only gravitationally, so the collisionless dark matter fluid can be 
represented by a set of discrete point particles.

Gravitational, long-range interactions are peculiar because:
(i) they are inherently negative specific heat systems (i.e., taking 
energy away from it heats it up; see, e.g., \citealt{lyn77,pad90,lyn01}),
or, equivalently, internal temperature gradients are enhanced 
instead of being erased \citep{elz98};
(ii) the potential energy is a {\it superextensive} quantity, which 
leads to an intrinsic instability in the case of virialized
systems when their mass is increased \citep[e.g., ][]{heg03}.
The difficulty of characterizing the dynamical 
evolution and final equilibrium state of collisionless 
systems is deepened by the fact that the classical statistical
mechanics and thermodynamics have been little developed for gravitational
problems due to complications posed by the long-range, unshielded nature 
of the gravitational force \citep[e.g., ][]{pad90,pad05}.
Here we focus on also peculiar, but largely overlooked property of gravitational
N-body systems, the {\it Energy Ranking Preservation} phenomenon
(hereon, ERP). The ERP was first noted by Quinn and Zurek \citep{qui88};
it reemerged later in the literature, as a subsidiary result,
in different contexts such as the evolution of simulated galaxy mergers and
collapses \citep{kan93}, the numerical study of ``violent relaxation" processes
\cite[e.g., ][]{fun92a}, the applicability of the secondary infall
model to collapsing cosmological proto-halos \citep{zar96}, or the study of properties of dark matter halos \citep{hen99,dan03,dan06}. 

A gravitational N-body system displays ERP 
when the relative ordination or ranking of 
sufficiently large groups of particles with respect to their mean energy is 
preserved along their possibly intricate trajectories in the phase space. 
In the present context, ERP means that there is a ``coarse-grained"
sense in which the ranking relative to the mean one-particle, 
mechanical energy of given collections of elementary particles 
is strictly {\it not} violated during the
gravitational evolution of the system. As remarked
by Kandrup et al. \citep{kan93}, ``mesoscopic constraints"
seem to be operative at the level of collections of particles 
when a large N-body system evolves gravitationally towards equilibrium. 

The present work is part of a project first motivated 
from a study of the properties of 
elliptical galaxies in the context of the so-called ``fundamental 
plane" \citep{dan03}. We have subsequently analysed the
ERP in individual halos in a cosmological simulation and have found some
preliminary results on the behaviour of the coarse-grained level necessary
for the stablishment of the ERP \citep{dan06}. In the present paper,
we aim to address the ERP effect within a large-scale N-body 
cosmological simulation in a more general context (as opposed
to the restricted ``individual halos" approach of previous studies). 
We investigate the problem in a more global sense (all particles 
of the simulation, inside and outside halos will be considered)
and at different dynamical regimes and scales, 
namely, at the linear regime, where structures are not yet virialized, and at the
small-scale nonlinear regime, where particles probably are part of virialized or
quasi-virialized structures.  

In fact, the non-Hubble component of a galaxy velocity
through the universe (its peculiar velocity), due to the
acceleration caused by clumps in the matter distribution,
is an important quantity in cosmology \citep*[e.g.,][]{jus99,fuk01,lan02,fel03}.
Here we will be interested in studying the {\it Pairwise Velocity Distribution} (PVD) from the point of view of the ERP phenomenon. In other words, 
we analyse a cosmological N-body simulation  in the linear and nonlinear 
regimes as a function of different energy partitionings, under the perspective of
characterizing the behaviour of the underlying cosmic flow. 

The present paper is organized as follows: in Section 2, we present the methodology and illustrate the ERP. In Section 3, we analyse the behaviour of the PVD. In Section 4, we offer a bird's eye view in the literature to contextualize 
the ERP phenomenon within our current understanding of the
gravitational collapse and virialization mechanisms. 
In Section 5, we present the summary and final conclusions.

\section{Methodology and the ERP phenomenon}

In the present work, we used in our analysis a
$\Lambda$-CDM N-body simulation of the VIRGO  Consortium (the
$\Lambda$-CDM model is currently the most observationally favored
cosmological scenario, e.g. \citealt{spe03,spe06}). 

The simulation box has
$239.5$ Mpc/$h$ ($h=0.7$) of size, with $256^3$ particles, each with
mass of $6.86 \times 10^{10}$ M$_{\odot}$/$h$, therefore representing
approximately galaxy-sized objects. We have selected a random
fraction of the particles ($1$ out of $100$) in order that our analysis be
computationally feasible. We have also restricted our analysis to
the initial ($z=10$) and final ($z=0$) outputs of this cosmological
run. Particles of a  given initial model are sorted according
to their mechanical energies. The energy space is then partitioned into a
few (e.g. $10$) cells or bins of equal number of particles and for
each of these bins, the mean energy is calculated.  The mean energies
of these {\it same} collections of particles are then recalculated for
the final model and compared with their initial values. As we will show below,
generally these mean energy cells do not cross each other along the evolution
(ranking is preserved), although particles individually gain
or lose energy as the system virializes. 

\begin{figure}
 \includegraphics[width=7cm,height=7cm]{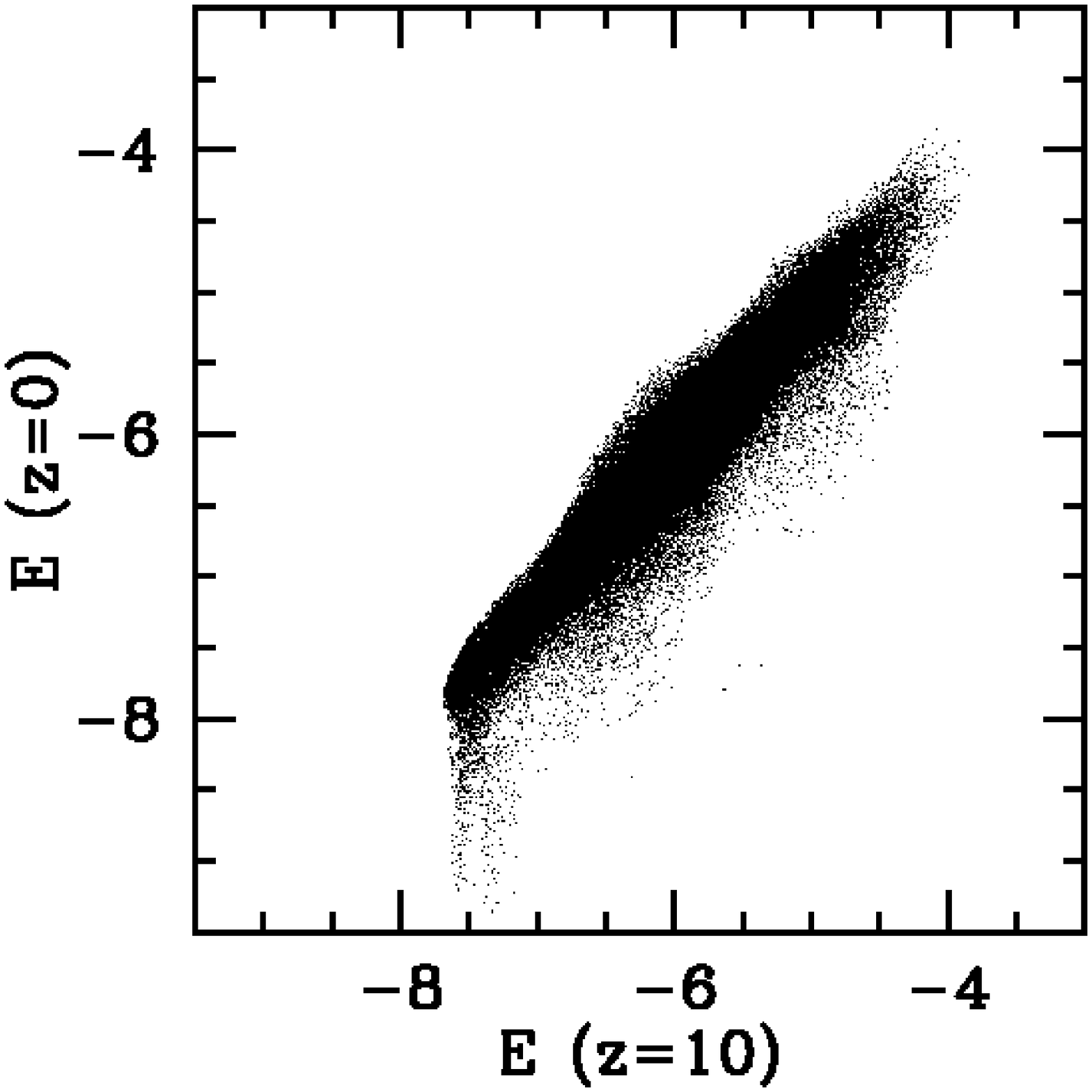}\\
 \includegraphics[width=7cm,height=7cm]{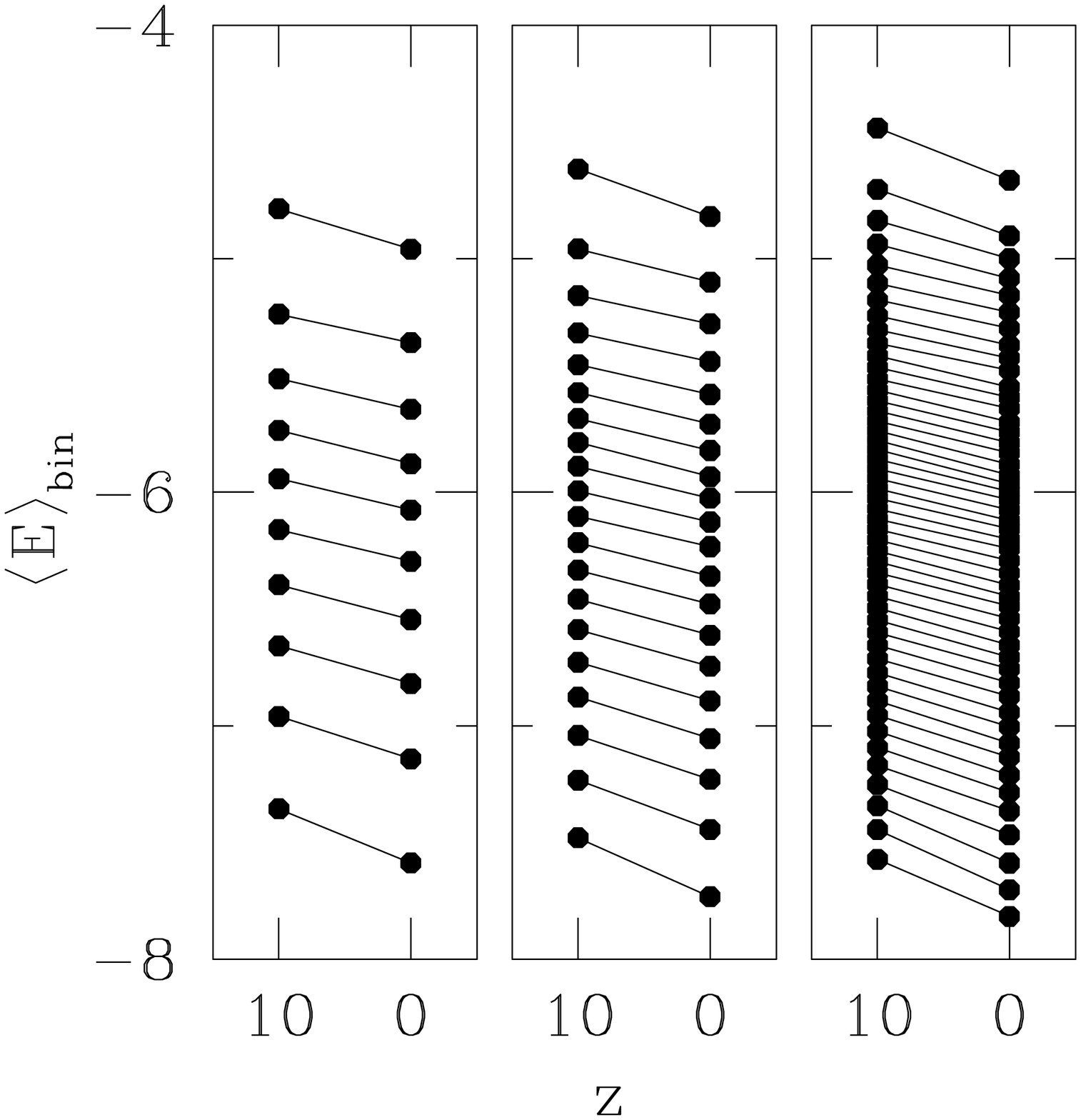}
\caption{{\it Top panel:} A scatter plot of the initial (redshift $z=10$) versus
final ($z=0$) energies of the particles in the cosmological simulation.
{\it Bottom panel:} The ERP effect: the ordering of the mean energies of
collections of particles are mesoscopically maintained for various levels
of energy partitionings ($N_{bins} = 10, 20$, and $50$; see text for 
details). Only the initial and final energy bins are shown here, connected
by line segments. \label{energy}}
 \end{figure}

In Fig. \ref{energy} we present two panels illustrating the global
characteristics of the energy space in the Virgo's $\Lambda$-CDM 
simulation. In the top
panel,  we show a scatter plot of the initial ($z=10$) versus final
($z=0$) energies of the particles in the simulation.  The mechanical
comoving one-particle energy of a
particle $i$ was calculated classically from $E_i = 1/2 m v_i^2 - 1/2 \sum_{j \neq
i} Gm^2/|{\bf x}_i - {\bf x}_j|$, with comoving position ${\bf x}_i$
and peculiar velocity $v_i$.  Units used were Mpc for length, Gyr for
time and $M_{\odot}$ for mass.  It represents the energy
with respect to the comoving reference frame (that is, the frame 
which moves with the cosmological expansion of the simulation box).
The energy associated to the dynamics of expansion of the cosmological 
box does not enter into the above computations.  

The straight $\sim 45^{o}$ line pattern exhibited in the top panel of 
Fig. \ref{energy} highlights the strong memory preservation of the initial
condition state along the evolution of the system in the energy space. 
We remark, however, that there is a significant spread around the
correlation, which indicates individual gain/loss of energy (`mixing')
during the evolution. At the bottom panel of Fig. \ref{energy}, 
we present the ERP effect for different
choices of energy partitioning ($N_{bins} = 10, 20$, and $50$; we have
fixed this choice of partitionings throughout this work).  The energy
cells do not cross, meaning that the most (least) energetic particles at the
initial state are still the most (least) energetic particles at the final
state. Note that the energy ordination is observed despite the fact that
particles individually have mixed in energy space. This unexpected orderly evolution contrasts with the highly complex structure
of the actual phase space of the system, which displays
sensitivity to changes in their initial conditions
which are characteristic of chaotic dynamical systems \citep{kan91,kan92,goo93,mer96,elz98,kan03}.
Hence, clearly, memory is preserved only in a 
coarse-grained, mesoscopic sense (top panel of Fig. \ref{energy}).

Notice also (c.f. top panel of Fig. \ref{energy})
that some of the most bound particles tend to
become at the end state even more gravitationally bound, as evidenced
by a tenuous `tail' at the bottom/left side of the scatter plot
(see also Fig. 13 of \citealt{qui88}, and Fig. 9 of \citealt{hen99}). 
The final mean coarse-grained energies are distributed in a 
smaller range than their initial counterparts. This would not be expected 
for energies evaluated at a fixed background. But in the present case, the 
energies are calculated with respect to the comoving frame (for further
details, see \citealt{dan06}).

\section{The Pairwise Velocity Distribution}

We will proceed our analysis considering the 
{\it pairwise velocity distribution} (PVD)
from an `energy space' perspective. 
Here we consider a pair of galaxies lying at a comoving separation vector
$\bf{R}$. We define the pairwise peculiar velocity as:

\begin{equation}
\delta v_{\parallel}(R) \equiv {1 \over R} (\bf{v_i} - \bf{v_j}) \cdot {\bf{R}} ,
\end{equation}
where $\bf{v_i}$ and $\bf{v_j}$ are the proper peculiar velocities  of
a given pair of galaxies in the system. The PVD is the distribution of
all particles $\delta v_{\parallel}(R)$ for a given scale $R$
probed. The importance of this quantity cannot be underestimated.
Notice that, separated at cosmological  distances, a given galaxy pair will
be approximately following the uniform Hubble flow.  In other words,
it is expected that the ratio $\delta v_{\parallel}(R)/(HR)$,
where $H$ is the Hubble parameter, tends to zero in this case.
On the other hand, objects at much smaller scales evolving under  
nonlinear regimes will have their dynamics supersede the 
Hubble flow \footnote{Davis \& Peebles \citep{dav77} 
have argued that for an isolated system in
virial equilibrium,  and therefore highly nonlinear, the physical
(proper) separation of particle pairs within the system is expected to
remain constant on average (the so-called ``stable clustering" hypothesis). 
In the universe,  however, no system is
exactly isolated, and  continuous merging and accretion are expected to
occur in a hierarchical fashion. Despite this complication, the stable clustering
hypothesis has been used as a model to the power spectrum of density fluctuations
encompassing nonlinear structures \citep{ham91}, and has been
extensively investigated for halos resulting from cosmological
simulations \citep[e.g., ][]{jin01}.}.

Hence, the properties of the PVD, evaluated at different comoving
scales, are intimately related to linear/nonlinear regimes. 
Our major concern here is to describe the 
behaviour of the PVD considering the fact that there is an
ordering preservation of the mesoscopic particle energy cells
during the cosmological evolution. How does the nonlinear
evolution of the pairwise motions behave under
such an energy constraint? 

\begin{figure}
\centering
 \includegraphics[width=7cm,height=8cm]{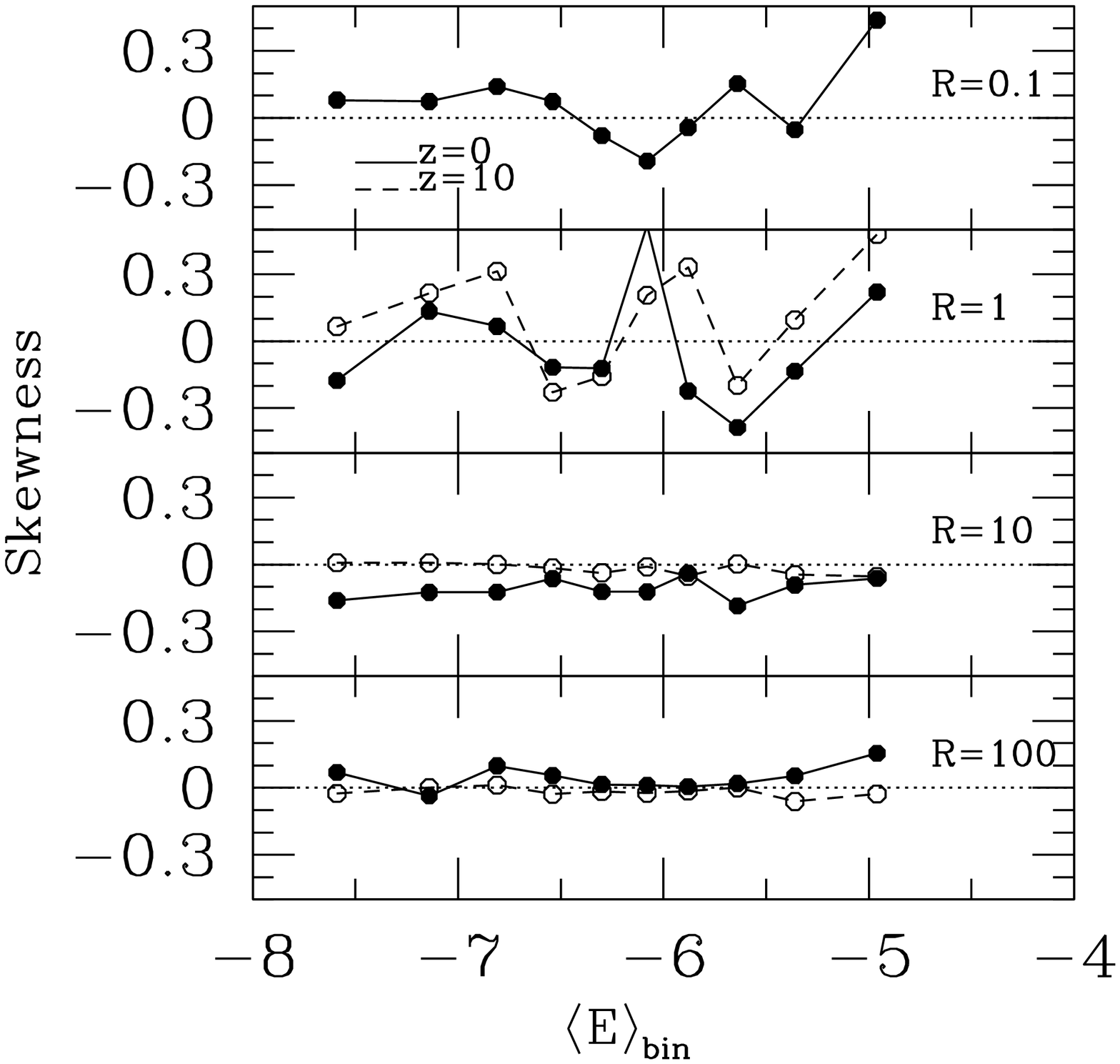}\\
 \includegraphics[width=7cm,height=8cm]{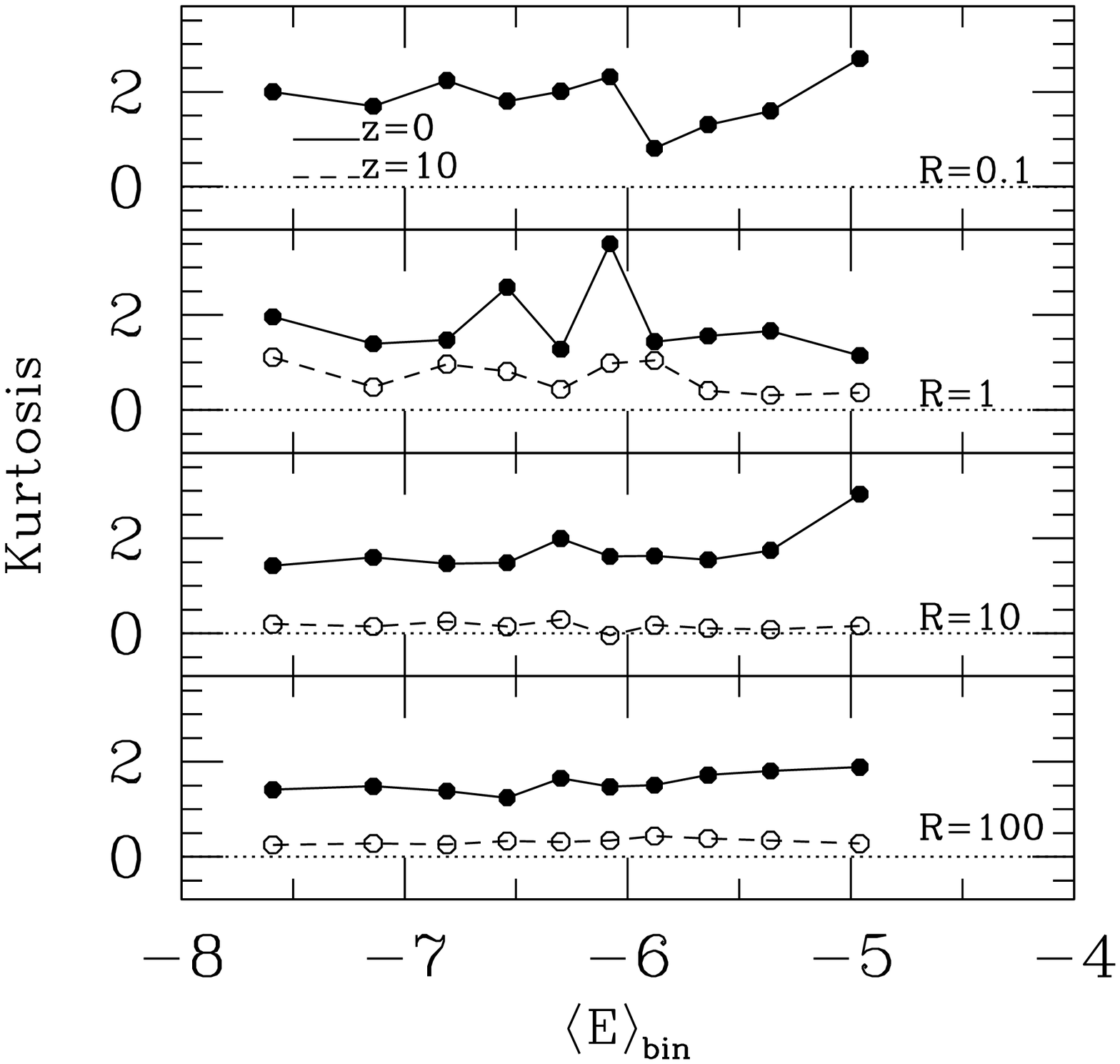}
 \caption{Skewness (top figure) and kurtosis (bottom figure)
of the PVDs obtained separately for
every energy cell as a function of the mean energy of the cell
for $z=0$ (solid lines) and $z=10$ (dashed lines). 
$N_{bins}$ fixed to $10$ in all these plots. Dotted lines indicate 
the expectations for a perfectly gaussian distribution. 
The absence of data for $z=10$  in the first 
top panels of the figures is due to lack of statistics in this scale ($R=0.1$).
\label{SK}}
 \end{figure}

Fig. \ref{SK} is an illustration ($N_{bins}$ fixed to $10$) of the typical
results of Skewness and Kurtosis\footnote{The Kurtosis, 
as usually defined, was decreased by $3$.} of the PVDs for every energy cell
separately. All PVDs at $z=0$ show significant departures from gaussianity.
Although this is a behaviour well known in {\it unconstrained} PVDs, and modelled analytically in previous works \citep[e.g., ][]{dia96, she96, set98}, it is not
obvious why this feature should remain in a energy-constrained PVD.
In fluid turbulence, for instance, energy-constrained PVDs recover gaussianity \citep{nae98}, indicating that the power-law tails apparent in the 
velocity histograms stem directly from the spatially intermittent nature of the underlying energy cascade that drives the system \citep{fri95}. 
This is clearly not the case here. 

More unexpectedly in Fig. \ref{SK}, 
is the evident correlation between early and late 
epoch statistics, for certain scales (R=1 and 10 Mpc). Note that scales below 
$10$ Mpc are precisely those that probe the nonlinear regime, where peculiar 
velocities dominate the Hubble flow and particles are already part of virialized 
or quasi-virialized clusters. On the other hand, 
on larger scales ($R \geq 100$ Mpc), where there is little 
discernible structure in cosmological box and the mass distribution appears homogeneous and isotropic, there is no evidence of memory 
on the skewness and kurtosis of the energy-constrained PVDs. 

\begin{figure}
\centering
 \includegraphics[width=4.17cm,height=5.2cm]{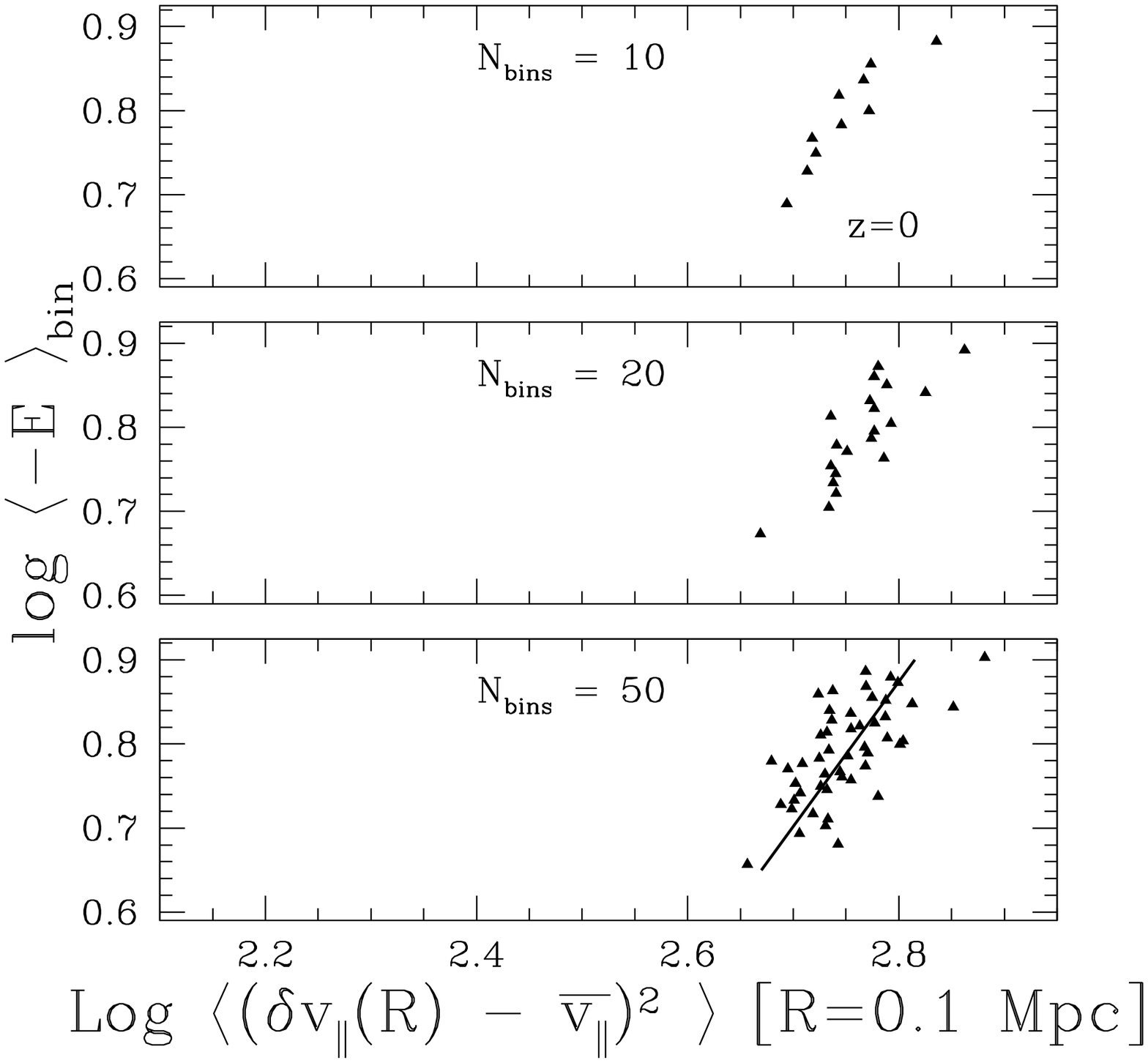}
 \includegraphics[width=4.17cm,height=5.2cm]{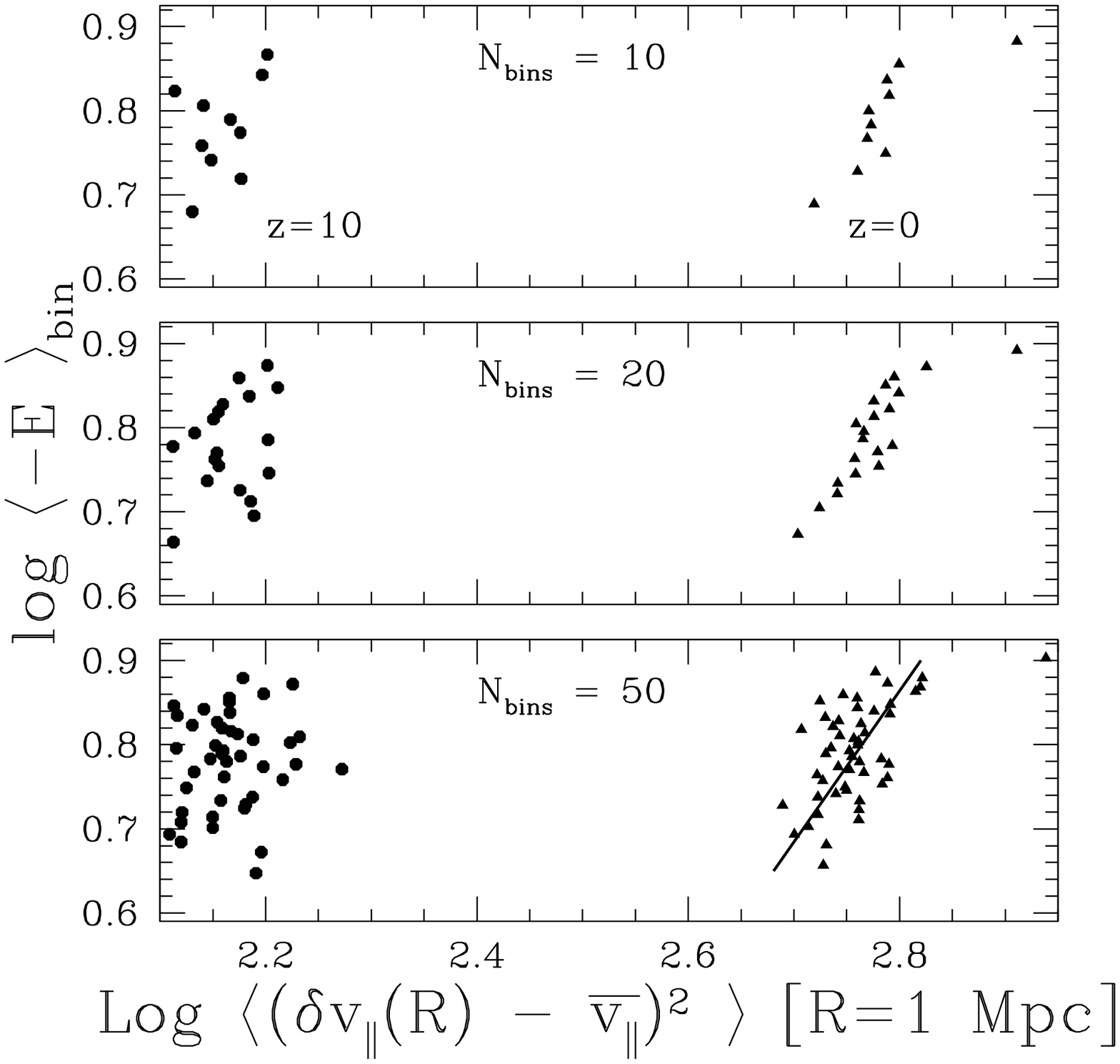}\\
 \includegraphics[width=4.17cm,height=5.2cm]{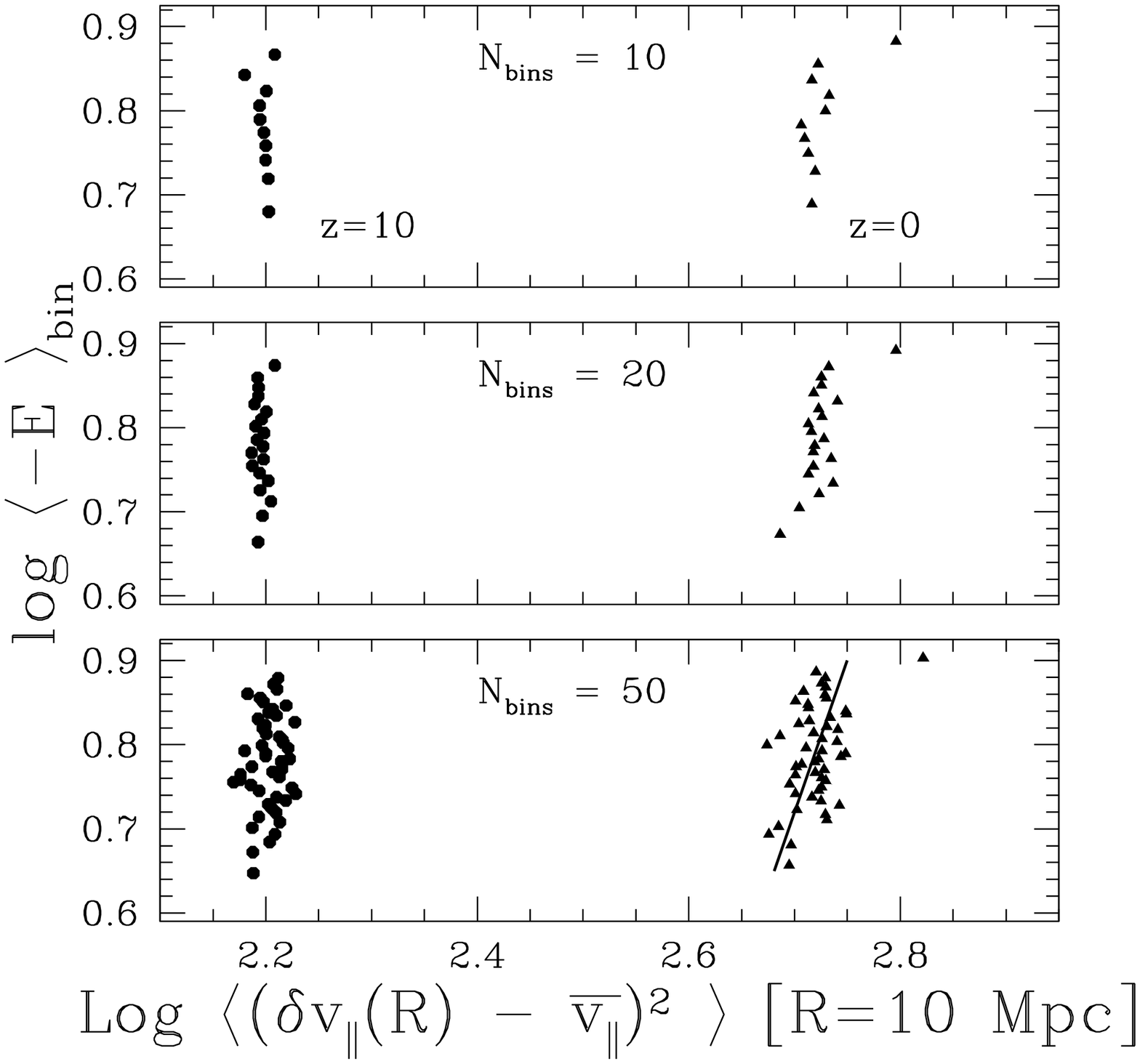}
 \includegraphics[width=4.17cm,height=5.2cm]{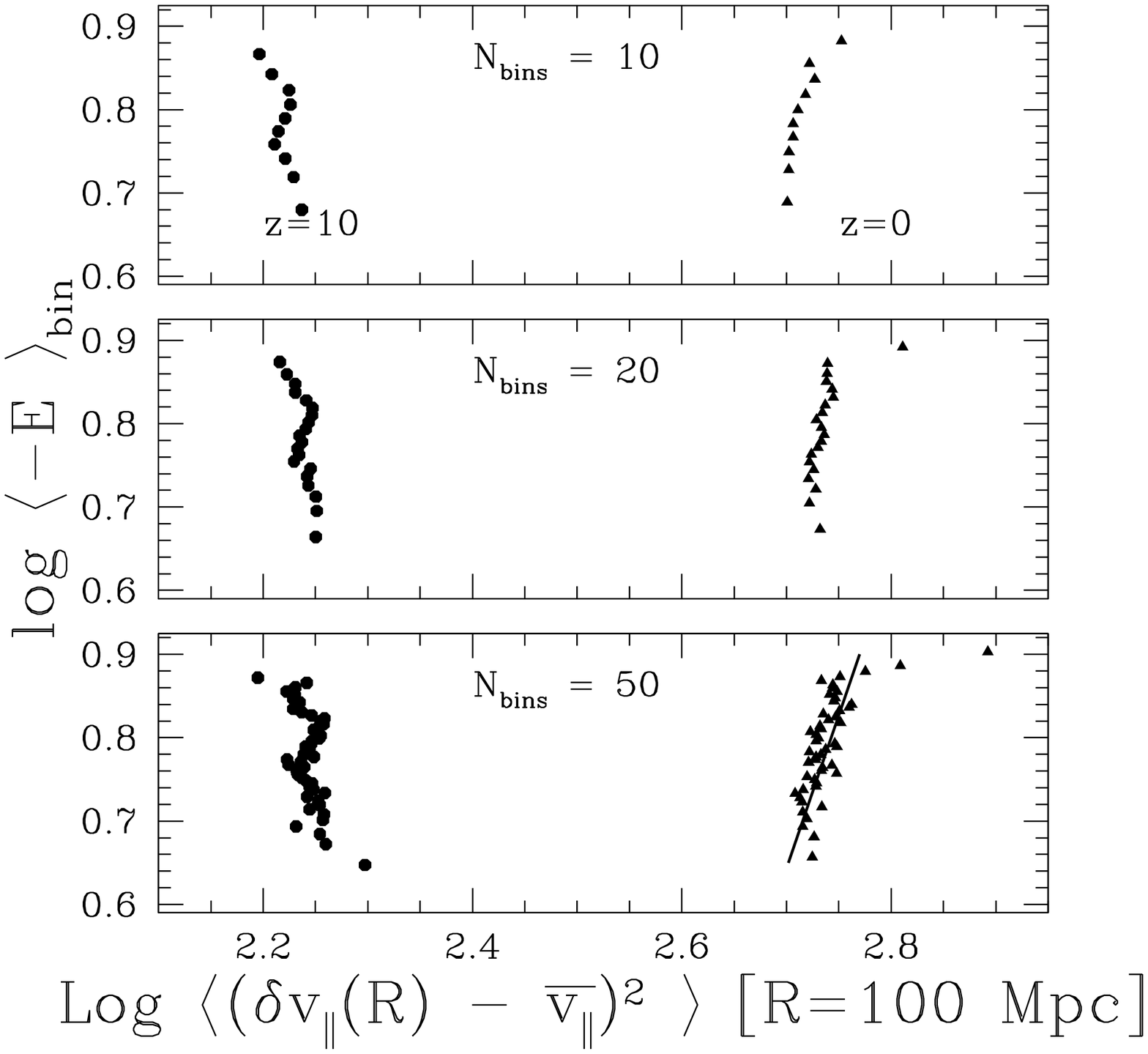}
 \caption{The coarse-grained mean energies as a function of the 
related dispersion of the PVDs (see text for details).  This mosaic
shows the results for every scale $R$ probed by the estimator;
{\it top panels:} results for an energy partition of $N_{bins} = 10$;
{\it intermediate panels:} $N_{bins} = 20$;
{\it lower panels:} $N_{bins} = 50$. {\it Circles:}
results for the initial state ($z=10$); {\it triangles:}
results for the final state ($z=0$). 
The absence of points in the first upper panel 
is due to lack of statistics in this scale for $z=10$.
Straight lines are fits to the data points, at $z=0$ and $N_{bins}=50$,
disconsidering the points corresponding to the boundest energy cells. The
corresponding slopes ($\alpha$) for these fits (according to scale $R$)  are:
$(\alpha, R) = (1.130,0.1), (1.562, 1), (3.215,10), (4.167,100)$.
\label{edisp}}
\end{figure}

In Fig. \ref{edisp}, we plot the dispersion $\langle (\delta v_{\parallel} (R) -
\bar{v_{\parallel}})^2 \rangle$) of the PVD as a function of the mean
energy cell for the initial ($z=10$, to the left, circles) and
final ($z=0$, to the right, triangles) states. 
At this point let us review what has actually been measured and its
meaning: (i) we measure the radial-component, peculiar, velocity difference 
of each pair of particles, for each mesoscopic
energy bin, to obtain the energy-constrained
PVD (Eq. \ref{energy}). This results in a distribution of velocity differences 
that is well-known to deviate from gaussianity in the energy-unconstrained 
case \citep{dia96, she96, set98}; and (ii) 
we measure the dispersion of the resulting energy-constrained PVD. 
Notice then that what is being considered it is {\it not} the velocity dispersion 
of the particles of a given energy cell. It is the dispersion of the PVD for 
each energy cell. In other words, the former and the later are different physical measures.

Overall, we observe that the more (negative) energetic the cell, the larger the 
corresponding dispersion. Moreover, the relation between dispersion 
and energy follows a power-law in which the characteristic
exponent increases with the probed scale (c.f. Fig. \ref{edisp} 
and corresponding legend). Notice, however, that probing smaller scales does 
{\it not} necessarily mean that one is also probing more (negative) 
energetic cells of central halo regions, which possess higher velocity 
dispersions, since, as already mentioned, the {\it velocity dispersion
and the dispersion of the PVD are different physical measures}.

In the case of less (negative) energetic cells (fix the triangle dots
mostly at the bottom of each fitted line in Fig. \ref{edisp}), 
for redshift $z=0$, moving to smaller scales in fact {\it decreases} 
the dispersion of the PVD (the characteristic exponent decreases for smaller
scales). Clearly, these less (negative) energetic cells, even at small
scales, are probing, most of the time, 
not yet virialized motions, still in the linear regime.
This result is in agreement with those of \cite{dia96}, where
it is found that unrelaxed systems tend to have more centrally peaked
PVD than gaussian distributions. Also, some of the particles composing these
less (negative) energetic cells, at small scales, can also make part
of the population of the outermost parts of dark matter halos,
which are not yet virialized. In any case, all energy cells follow 
the same simple scaling law, regardless of the linear or nonlinear 
clustering regime. 

\section{Virialization and the ERP phenomenon}

Up to this point we have shown that the ERP is a phenomenon
that occurs embodying all particles of a $\Lambda$-CDM 
cosmological simulation, regardless of the virialization condition 
of the immediate regions to which the particles are associated with.
Also, it is reasonable to consider that the energy-constrained PVDs
scaling laws found in the present work are a reflection of a
remarkably smooth dynamical evolution of the large-scale motions resulting from the
clustering process taking place in the simulation. In this section, we 
allow for a brief  digression and review the relevant literature on the
virialization of dissipationless gravitational systems
in order to bring the ERP effect into this context.

\subsection{Dissipationless gravitational systems}

In a first approximation, if the dissipation and evolution of the 
baryonic component of a gravitational system can be 
disregarded, the problem of virialization can be studied by the means
of numerical experiments that attempt to reproduce the purely 
gravitational evolution of a set of $N$ point masses\footnote
{Dissipational effects tend to contribute to the formation
of deep central potentials \citep[e.g., ][]{car86}.}. 
Well before such experiments could properly
be performed, the theoretical work by Chandrasekhar in 1943 \citep[viz., ][]{bin87}
already indicated the existence of different dynamical regimes, namely:
if ${(8 \ln N) / N} << 1$, the system is collisionless in the sense that
each particle responds to the average gravitational field of the system;
otherwise, the system is collisional or marginally collisional. 

In the collisional case, local fluctuations 
or granularities of the gravitational potential are important to the overall
dynamics. The accumulative effect of the energy exchange between 
stars as they deflect each other during an encounter results that,
at the gravitational two-body relaxation time-scale, the ``memory" of the
initial orbits of the stars have been erased.
Objects like globular cluster are considered examples of such systems.
In the collisionless case, however, the corresponding
relaxation time-scale is longer than the age of the universe, 
hence the system cannot reach a thermal-equilibrium state from binary
collisions. Therefore, it can only relax to a quasi-equilibrium state
through some other mechanism, which also must operate rapidly enough 
(see detailed discussion and references in the next subsection).
Elliptical galaxies and dark matter halos are believed to be examples of
such systems. 

\subsection{Brief review on virialization of gravitational systems}

The first attempts to provide some statistical/thermodynamical 
description of gravitational stellar dynamics and collisionless relaxation
started in the 1960s with the pioneering work of Antonov 
(see, e.g., \citealt{lyn01} and references therein), 
Ogorodnikov \citep{ogo65}, Saslaw \citep{sas68, sas69},
and Lynden-Bell \citep{lyn67}. Lynden-Bell attempted to answer the problem 
of how a collisionless system would be able rapidly to relax to a 
quasi-equilibrium state, since the equipartition
of energy cannot proceed through two-body encounters. 
The resulting theory (``violent relaxation")
stated that {\it if} rapid fluctuations in the mean gravitational
potential energy field ($\langle \phi \rangle$) of a system far from
equilibrium persisted long enough before damping, these oscillations would
promote the mechanism for changing the energy per unit mass ($E_{\star}$) of 
each star in the system ($dE_{\star}/dt = \partial \langle \phi \rangle
\partial t$). Notice that $E_{\star}$ can change by a large amount under 
this hypothesis, that is, $|\Delta E_{\star}| \sim |E_{\star}|$, so that
the effect would be as significant as in the two-body relaxation case.

Soon after Lynden-Bell's publication, several one-dimensional N-body
numerical experiments were performed to test the outcome of gravitational collapse
against the predictions of ``violent relaxation", but the results were
mostly inconclusive (e.g., 
\citealt{hoh67,hoh68,hen68,gol69,cup69,lec71};
see also more recent developments along
these lines in \citealt{tan87,min90}). These simulations
generally started with a given initial volume of phase-space
uniformly occupied by all particles of the system, and the resulting
end state object usually presented a degenerated core and a distinct
halo population containing most of the mass (see also \citealt{bov70,got73}),
which could only be partially fit to Lynden-Bell's distribution.
These results indicated that ``violent relaxation" would probably
be an incomplete process, in the sense that the fluctuations of the 
gravitational potential would damp too rapidly and not all phase-space
cells dynamically accessible to the system would end up occupied.
Consequently, the final equilibrium product generally 
resulted in a partially relaxed object with a stationary potential 
(except for small statistical fluctuations) that would not necessarily 
represent the configuration that maximized the total entropy of the system.

Notice that the general idea behind most of the approaches to derive a theory
of collisionless relaxation is to assume that the final 
equilibrium state will be the most probable state
of the system. Such a state could be predicted from first 
principles by maximizing the entropy of the system under the appropriate constraints
(usually, the total mass and energy), taking into account the
fact that the phase-space density must evolve under the collisionless 
Boltzmann equation. The standard result turns out to be, however,
that the entropy is extremized if and only if the distribution function
is of the isothermal sphere. But the energy and mass of
an isothermal sphere diverge for large radii, so this leads to a contradiction
with the initial assumptions of the problem. The general conclusion is that
{\it no} DF compatible with fixed mass and energy maximize 
the entropy: arbitrarily large entropy configurations can 
be obtained by an arrangement of the particles of the system 
\citep[e.g., ][]{bin87}\footnote{This fact lead to
several investigations on the properties of {\it truncated} isothermal 
spheres, i.e., constrained to a finite spherical shell
and truncated at a given energy. Although such a procedure
seems too arbitrary, it can be formally justified
observing that the external parts of a gravitational N-body
system behave just like the boundary 
conditions artificially introduced by a fixed shell.
Such an approach has been proved useful to the physical modelling of the
problem, since the type of shell introduced (and consequently
the type of physical quantity that is exchanged with the external
reservatory) describes the system as a member of a specific kind of 
{\it ensemble} \citep[e.g., ][]{pad90}.}. 
For instance, the entropy can increase without bound by 
confining a small fraction of the mass in a compact core and
arranging the remainder in a diffuse halo. Although the existence 
of such a core-halo configuration is a natural consequence of the relaxation
of collisional systems, like globular clusters, it does not seem 
to be a physically accessible one to collisionless 
systems, since elliptical galaxies and dark matter halos are not observed
to have such structures.
 
Lynden-Bell was aware that in real systems ``violent relaxation" would
probably not proceed to completion (see section 6 of his 1967 paper).
Hence the theory would only represent a qualitative idea that needed 
to be complemented by other mechanisms  and/or superseded by a more robust 
theory. This line of thought was strengthened more and more
from different fronts, in particular from theoretical and
observational studies of the scaling laws in elliptical galaxies
\citep[e.g., ][]{bin82,tre86,whi87,ric88,hjo91,hjo93,hjo95,tre05a,tre05b}, 
and from more elaborate 3D N-body
simulations \citep[e.g., ][]{van82,mcg84,may84,vil84,agu88,pal90,fun92b}.
These simulations showed that not
all initial conditions would lead to the same final end-state;
in special, only clumpy, ``cold" initial states would preferentially
lead to objects more closely resembling the profiles of elliptical
galaxies. 

Further extensions and/or elaborations over Lynden-Bell's idea
\citep[e.g., ][]{sev80,gur86,spe92,kul97}, as well as other developments towards 
a more consistent theory of gravitational relaxation
\citep[e.g., ][]{shu78,wie88,zie89,zie90,cha96,nak00}, were subsequently carried 
out. The main difference 
between several of these approaches lies in the adopted definition 
of entropy and the constraints under which it is maximized. 
Even though there is now a large body of work
towards a theory of the gravitational relaxation phenomenon, it
remains a challenging endeavour, and in fact a deeper evaluation of
some of these theories lead to severe
inconsistencies \citep[viz., ][]{ara05a,ara05b}\footnote
{Notice at this point that another physical mechanism 
that is known to contribute to the
collisionless relaxation of gravitational systems is ``phase
mixing", in which the phase-space microscopic cells are continually
stretched into thinner and thinner filaments as relaxation proceeds, until
the coarse-grained phase-space density no longer changes, and the
system is said to have reached a coarse-grained steady-state configuration
\citep{bin87}. However, ``phase mixing"  and ``violent relaxation"
(whatever the given potentials admit or not the coexistence 
of regular and chaotic orbits) are in fact special cases of a more
general relaxation process of collisionless gravitational systems
\citep{zie94,kan98}.}.

Parallel to the developments is stellar dynamics, 
N-body cosmological simulations of hierarchically clustering model
universes started to address, at increasingly higher spatial and temporal
resolutions, the issue of the dynamical and structural properties of
virialized dark matter halos. For instance, the importance 
of mergers of smaller subsystems -- as opposed to simple collapses -- 
to the formation of elliptical galaxies and  
dark matter halos started to become clear important issues
 \citep[e.g., ][]{fre85,zhan02}. The main result however
is the existence of a universal halo profile \citep{nav97,nav04}, 
to which an explanation
in terms of the dynamical evolution and expected final equilibrium state
is still being searched for \citep[e.g., ][]{deh05,aus05,bar05,lu05}. 
Notice that, as previously mentioned, simulations of isolated collapses 
(and mergers) tend to produce different end states depending on the
initial conditions. So the apparent universality of dark matter
halo profiles in cosmological simulations 
could be interpreted as the fact that the underlying relaxation process 
in such scenarios produce systems in a state close 
to the most probable one, as derived according to 
some more fundamental relaxation principle, yet to be uncovered.

\subsection{How does the ERP phenomenon fit in?}

Clearly, ``violent relaxation" is 
routinely observed in N-body experiments,
but it rapidly damps out at about the free-fall timescale of the system.
Gravitational potential fluctuations can be violent enough to
mix the one-particle phase-space structure, but not so violent 
as to remove all ``memory" of the initial coarse-grained, mean
particle energy ordering. 

Dissipationless relaxation must therefore be described as
a process in which collections of particles closely spaced in energy space can
efficiently mix in phase space, but underlying such a mixing,
there is a restriction to the exchange of energy, 
which must proceed within some upper and lower bounds.
Such bounds can only change linearly and orderly with respect to
all other energy cells of the system, or at least must be restored
subsequently to the major potential fluctuations.
This suggests that a line of approach to the EPR effect
that might prove fruitful is a formulation based on a 
maximum entropy principle in which a further constraint,
in addition to the total mass and energy,
must be imposed, namely, a ``mesoscopic constraint",
operative only at the level of a coarse-grained distribution
function \citep{kan93}.

Some clues on how this theory must be accomplished, at our present 
stage of understanding, are outlined below:

\begin{itemize}
\item{The development of a physically well-motivated, incomplete 
treatment of the ``violent relaxation" process will be an indispensable
step to the formulation \citep[see, e.g., ][]{hjo91}.}
\item{The underlying ``mesoscopic" constraint must hold even under
large fluctuations of the gravitational potential field \citep{kan93}.}
\item{Some heuristic inputs must be taken into account, which we
collect from present and previous work, namely:}
\subitem{(i) evidences that the ERP depends
on halo mass, in the sense that more massive halos show more
rank preservation than less massive ones \citep{dan06}.}
\subitem{(ii) some reasonable validity of Arnold's theorem when
applied to individual halos, showing that the ``mesoscopic"
constraint acts at the level where collections of particles
behave dynamically as an individual particle with a characteristic 
frequency (or alternatively, energy) in the mean potential
field of the halo \citep{dan06}.}
\subitem{(iii) the fact that cosmic flows evolve in an orderly sense, 
when tracked from their coarse-grained energy cells,  even when 
nonlinearities are already developed (Figure \ref{edisp} of present
paper).}
\subitem{(iv) other ``memory"-like effects, like those evident in
higher-order statistics of the PVDs, for scales $R \leq 10$ Mpc
(Figure \ref{SK} of present paper).}
\subitem{(v) other scaling correlations associated with the ERP
phenomenon, namely, that simulations of galactic mergers produce
``fundamental-plane"-like objects and some energy rank {\it violation}; 
whereas simple collapses are homologous systems and show {\it clear} ERP
\citep{dan03}.}
\end{itemize}

\section{Conclusion}

Summarizing, in this paper we have analyzed the behaviour 
of a cosmological N-body simulation in the linear and nonlinear 
regimes as a function of different energy partitionings. We have
found evidences that ERP effect, already observed in different, 
more restricted contexts, is also present in a large-scale, realistic 
cosmological N-body simulation from the VIRGO Consortium. We have observed 
that ERP-like ``memory" effects are also evident in
higher-order statistics of the PVDs, for scales $R \leq 10$ Mpc. 
Finally, we have noticed that there is a general scaling law 
relating the energy cells to the energy-constrained PVD velocity 
dispersion that appears to hold regardless  the dynamical regime being probed.

The expansion of the cosmological background brings more complexity to 
the virialization problem when compared with the ``simpler" studies of statistical 
mechanics of isolated gravitating systems \citep{pad98}. 
There are, however, common effects that persist in both scenarios,
like the ERP phenomenon here investigated. Although
several insights can be obtained from the bulk of current
knowledge on the nonlinear gravitational evolution, a complete
theory is yet to be derived \citep{pad05}. For instance, when viewed from 
the energy space perspective, the cosmic motions and clusterings,
at several stages of development (from the linear  throughout the
nonlinear phase), do display a certain degree of simplicity and an unexpected 
level of order. The ERP phenomenon
is certainly an important ingredient to the outstanding open problem
of the nature of the collisionless relaxation process.

\section*{Acknowledgments}
The simulations in this paper were carried out by the Virgo 
Supercomputing Consortium using computers based at Computing Centre 
of the Max-Planck Society in Garching and at the Edinburgh Parallel 
Computing Centre. The data are publicly available at 
{\tt www.mpa-garching.mpg.de/NumCos}. We thank Dr. Andr\'e L. B.
Ribeiro for useful discussions. This work was partially supported
by Fapesp and CNPq-Brazil.

\bibliographystyle{mn2e}
\bibliography{DantasAndRamos06MNRAS}

\end{document}